# When Celebrities Endorse Politicians: Analyzing the Behavior of Celebrity Followers in the 2016 U.S. Presidential Election

Yu Wang
*University of Rochester*
ywang176@ur.rochester.edu

Jiebo Luo
*University of Rochester*
jluo@cs.rochester.edu

## Abstract

Thanks to their name recognition and popularity, celebrities play an important role in American politics. Celebrity endorsements could add to the momentum of a politician's campaign and win the candidate extensive media coverage. There is one caveat though: the political preference of celebrity followers might differ from that of the celebrity. In this paper we explore that possibility. By carefully studying six prominent endorsements to the leading presidential candidates in the 2016 U.S. presidential election and statistically modeling Twitter *follow* behavior, we show (1) followers of all the celebrities with the exception of Lady Gaga are more likely to follow a large number of candidates and (2) the opinion of celebrity followers could systematically differ from that of the celebrity. Our methodology can be generalized to the study of such events as NBA players' refusing to visit the White House and pop singers' meeting with Dalai Lama.

## Introduction

Celebrities have long been known to play an important role in American politics [1], [11], [34], [37]. They serve as "information shortcuts" of the candidates whom they either explicitly support or have formallly endorsed [36], [37]. Taking advantage of their own name recognition, celebrities can make public performances on behalf of the candidates, they can campaign alongside the candidates, they can attend public conferences to voice their support, and more recently they can tweet out their endorsement to their millions of followers.

A few recent events, however, have raised the question whether celebrities and celebrity followers necessarily share the same political opinion. One such event is Meryl Streep's attack on Donald Trump and the subsequent criticisms from Streep's fellow celebrities, such as Travis Tritt. Another is Jennifer Holliday's decision not to perform at Trump's inauguration due to opposition from her fans [33]. As a third example, because of pressure from employees, Uber C.E.O quit President Trumps economic advisory council [17]. These events invariably suggest that the political opinion of the followers could differ from that of the celebrity.

In this paper, we explore this possibility through investigating an exhaustive data set that has recorded the revealed preferences of 15.5 million individuals on Twitter in April 2016. We apply negative binomial regression to study the Twitter *follow* behavior of celebrity followers and we employ the multinomial logistic regression to examine whether among these 15.5 million individuals, followers of a certain celebrity share the political preference of the celebrity.

The celebrities covered in our study are Lebron James, Lady Gaga, who have endorsed Hillary Clinton, Willie Robertson, Toby Keith, who have endorsed Donald Trump, Cornel West and Susan Sarandon, who have endorsed Bernie Sanders (Figure 1). All the celebrities satisfy the following two criteria: a). they have endorsed the respective candidate and b) they have a large number of followers on Twitter.

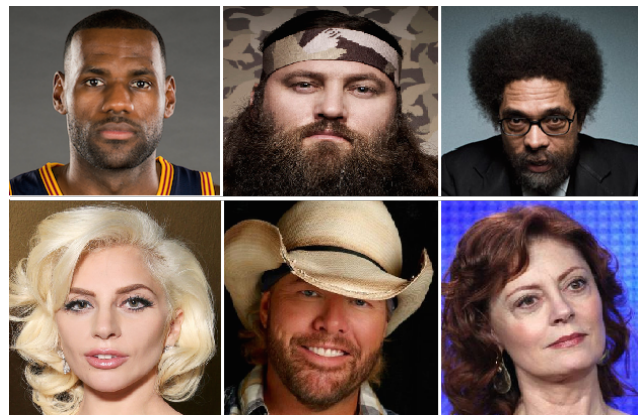

Figure 1. Left column (Clinton endorsers): Lebron James and Lady Gaga; center column (Trump endorsers): Willie Robertson and Toby Keith; Right column (Sanders endorsers): Cornel West and Susan Sarandon.

The contributions of our paper are as follows:
- we make a comprehensive analysis of the topology of the Twitter sphere during the 2016 presidential election.
- we analyze the behavior of celebrity followers with respect to following presidential candidates.
- we study the alignment and misalignment of political preferences between celebrities and celebrity followers.

## Related Literature

Our paper builds on previous literature on electoral studies, data mining and computer vision.

While the role of celebrities in American politics could be traced back to the early 20th century [29], research on the effects of celebrity endorsements on election outcomes only started recently, particularly since Oprah Winfrey's endorsement of Barack Obama in 2008 [11], [34]. Relationship between the credibility of the endorsers and the effect of the endorsements is analyzed in [27], [30]. Bernie Sanders also attributes part of his campaign success to the celebrity support that he has received [37]. Instead of focusing on examining the causal relationship between celebrity endorsements and voting outcomes, our study explores the alignment and misalignment of the revealed preferences of celebrities and their followers.

Existing work has also studied the increasing polarization of American politics at both the elite level [14], [25] and the mass level [6], [7]. Druckman *et al.*, in particular, study how elite partisan polarization affects public opinion formation and find that party polarization decreases the impact of substantive information [9]. Social clustering, on the other hand, is analyzed in [2], [26]. In our work, we contribute to analyzing political polarization at the public level on Twitter.

Gender plays an important role in the forming and dissolving of relationships [5], in online behavior [32] and in political voting [3], [8], [20], [40]. One common observation is that women tend to vote for women, which is usually referred to as gender affinity effect. In this paper we will control for gender effects when analyzing the number of presidential candidates that an individual chooses to follow and on which party that one chooses to follow.

Given the importance of gender in real applications, a large number of studies have attempted to classify gender based on user names [28], [31], tweets, screen name and description [4] and friends [43]. Following this line of research, our study will take advantage of information from both user names and user-provided descriptions.

Recent advances in computer vision [15], [21], [38], [39], on the other hand, have made object detection and classification increasingly accurate. In particular, face detection and gender classification [10], [18], [23] have both achieved very high accuracy, largely thanks to the adoption of deep learning [22] and the availability of large datasets [16], [19], [35]. Our paper extracts gender-related information based on Twitter profile images and is related to gender classification using facial features [23], [31], [40], [41].

## Data and Methods

Our dataset consists of three components, all of which come from Twitter and are collected using Twitter REST APIs[1]. The first component consists of the followers' Twitter

[1] https://dev.twitter.com/rest/public.

Table I
The Number of Followers (April, 2016)

| Candidate | # Followers | Candidate | # Followers |
|---|---|---|---|
| Chafee (D) | 23,282 | Clinton (D) | 5,855,286 |
| O'Malley (D) | 130,119 | Sanders (D) | 1,859,856 |
| Webb (D) | 25,731 | Bush (R) | 529,820 |
| Carson (R) | 1,248,240 | Christie (R) | 120,934 |
| Cruz (R) | 1,012,955 | Fiorina (R) | 672,863 |
| Kasich (R) | 266,534 | Huckabee (R) | 460,693 |
| Paul (R) | 841,663 | Rubio (R) | 1,329,098 |
| Trump (R) | 7,386,778 | Walker (R) | 226,282 |

Note: Sorted by party affiliation and alphabetically.

ID information for all the presidential candidates in April, 2016. This component is exhaustive in the sense that we have recorded all the followers' IDs. In total, there are 15,455,122 individuals following the 16 presidential candidates and some of them are following more than 1 candidate. We transform this component into a 15.5 million by 16 boolean matrix, with each row representing an individual and each column a presidential candidate. We report the summary statistics in Table I. It can be easily observed that Donald Trump and Marco Rubio have the largest numbers of followers among the Republican candidates and that Hillary Clinton and Bernie Sanders have the largest numbers of followers among the Democratic candidates.

The second component of our dataset has 1 million individuals, randomly sampled, with Python seed set to 11, from the first component. Based on these individuals, we extract user name, user-provided description, the starting year of using Twitter, social capital [40], and the profile image [41].

The third component comprises follower information of Lady Gaga, Lebron James, Willie Robertson, Toby Keith, Cornel West and Susan Sarandon, six celebrities all of whom have either explicitly or implicitly endorsed either Clinton or Trump or Sanders. These six celebrities constitute a significant presence among individuals who follow the presidential candidates: 15.9% of the individuals in the dataset follow Lebron James, 19.58% follow Lady Gaga, 0.9% follow Willie Robertson, 1.1% follow Toby Keith, 1.6% follow Cornel West and 1.2% follow Susan Sarandon. This data component then enables us to analyze the celebrity effect in a quantitative manner: whether individuals who follow these celebrities are also more likely to follow a large number of presidential candidates or a small number and whether the political preferences of the followers are the same as that of the celebrity.

We summarize the variables used in this work and their definitions in Table II.

### Gender classification

We employ three methods to extract information on gender. As in several prior studies [28], [31], we first compile a list of 800 names, based on appearance frequency on Twitter,

Table II
VARIABLE DEFINITIONS

| Name | Definition |
|---|---|
| **Independent variables:** | |
| Tweets | Count, number of tweets posted |
| Social Capital | Count, number of followers |
| Journalist | Binary, a journalist |
| Female | Binary, female by first name or image or description. |
| Lady Gaga | Binary, follow Lady Gaga |
| Lebron James | Binary, follow Lebron James |
| Willie Robertson | Binary, follow Willie Robertson |
| Toby Keith | Binary, follow Toby Keith |
| Cornel West | Binary, follow Cornel West |
| Susan Sarandon | Binary, follow Susan Sarandon |
| **Dependent variables:** | |
| # Candidates | Count, number of candidates that one follows |
| Democrat follower | Binary, follow Democrats only |
| Republican follower | Binary, follow Republicans only |
| Independent follower | Binary, follow both Democrats and Republicans |
| Bernie Sanders | Binary, follow Bernie Sanders |
| Hillary Clinton | Binary, follow Hillary Clinton |

Note: (1) Following previous studies [40], we define social capital on Twitter as the raw number of followers. (2) By construction, Democrat follower, Republican follower and Independent follower always sum up to 1.

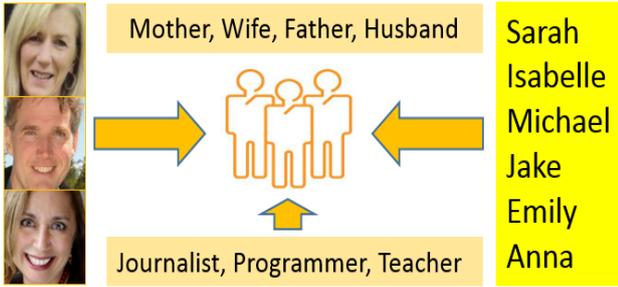

Figure 2. We use first names, profile images and family roles to identify gender, and we extract from self-descriptions individuals' occupations.

that are gender-revealing, such as Mike, Jake, Emily, Isabella and Sarah. This constitutes our first channel. We then use this list to classify individuals whose names are contained in this list. As one would expect, a large number of individuals can not be classified with this list.

Our second channel is the profile image. We train a convolutional neural network using 42,554 weakly labeled images, with a gender ratio of 1:1. These images come from Trump's and Clinton's followers. We infer their labels using the followers' names (channel 1). For validation, we use a manually labeled data set of 1,965 profile images for gender classification. The validation images come from Twitter as well so we can avoid the cross-domain problem. Moreover, they do not intersect with the training samples as they come exclusively from individuals who unfollowed Hillary Clinton before March 2016.

The architecture of our convolutional neural network is the same as in [42], and we are able to achieve an accuracy of 90.18%, which is adequate for our task (Table III).

Table III
SUMMARY STATISTICS OF CNN PERFORMANCE (GENDER)

| Architecture | Precision | Recall | F1 | Accuracy |
|---|---|---|---|---|
| 2CONV-1FC | 91.36 | 90.05 | 90.70 | 90.18 |

Third, we extract gender-revealing keywords from user-provided descriptions. These keywords are *papa, mama, mom, father, mother, wife* and *husband*.

We prioritize the first channel (first names) most and the third channel (self description) the least. Only when the more prioritized channels are missing do we use the less prioritized channels: first names > profile images > self descriptions. Based on this ranking, we are able to label 38.7% of the observations from first names, another 17.2% with profile images and 0.7% with self descriptions. In total, we are able to classify 56.6% of the 1 million individuals. We summarize the number of labeled individuals and the net contribution of each channel in Table IV.

Table IV
3-CHANNEL CLASSIFICATION OF GENDER

| Channel | First Name | Profile Image | Self Description |
|---|---|---|---|
| Priority | 1 | 2 | 3 |
| Identification | 387,148 | 304,278 | 30,786 |
| Contribution | 38.7% | 17.2% | 0.7% |

Note: Partly as a result of our priority ranking, the net contribution of profile images is significantly smaller than first names. The net contribution of self descriptions (3rd channel) is about 1 percent.

We summarize the gender ratio of each candidates' followers in Figure 3. It can be seen that Clinton has the highest female to male ratio, followed closely by Bernie Sanders. Rand Paul (R) and Jim Webb (D) on the other hand have the lowest female to male ratio. In general, the Democratic candidates mostly have a gender ratio close to or over 40%, while the Republican candidates tend to have a gender ratio well below 40%. Carly Fiorina, the only female candidate in the Republican party, is the only Republican to reach 40%.

*Negative binomial regression*

In order to understand how celebrity followers behave differently from non-followers, especially with respect to the number of presidential candidates that they follow, we apply the negative binomial regression [12] and link the number of candidates that one follows, which is count data, to the explanatory celebrity variable. In this regression, the conditional likelihood of the number of candidates that individual $j$ follows, $y_j$, is formulated as

$$f(y_j|v_j) = \frac{(v_j\mu_j)^{y_j}e^{-v_j\mu_j}}{\Gamma(y_j+1)}$$

where $\mu_j = exp(\mathbf{x_j}\boldsymbol{\beta})$ is the link function that connects our explanatory variables to the number of candidates that one chooses to follow and $v_i$ is a hidden variable with a Gamma($\frac{1}{\alpha}, \alpha$) distribution. After plugging in the explanatory variables, the unconditional log-likelihood function takes the form:

$$\ln L = \sum_{j=1}^{N}[ln(\Gamma(m+y_j)) - ln(\Gamma(y_j+1)) - ln(\Gamma(m))$$
$$+ mln(p_j) + y_j ln(1-p_j)]$$
$$p = 1/(1+\alpha\mu)$$
$$m = 1/\alpha$$
$$\mu = exp(\beta_0 + \beta_1 \text{Tweets Posted} + \beta_2 \text{Follower Count}$$
$$+ \beta_3 \text{Journalist} + \beta_4 \text{Year Fixed Effects}$$
$$+ \boldsymbol{\beta_5 \cdot \text{Female}} + \boldsymbol{\beta_6 \cdot \text{Celebrity}})$$

where $\alpha$ is the over-dispersion parameter and will be estimated as well.

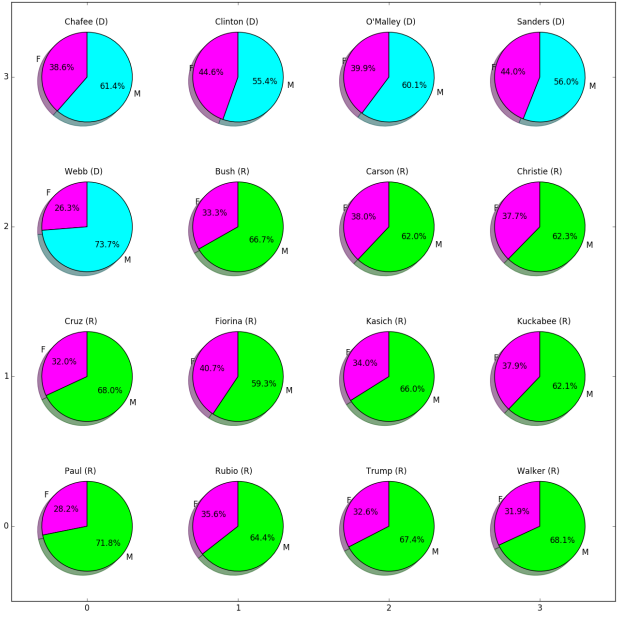

Figure 3. In percentage, the leading Democratic candidates have more female followers than the leading Republican candidates.

*Multinomial Logistic Regression*

Besides the number of candidates, we also explore the alignment and misalignment of political preferences between celebrities and celebrity followers. For this purpose, we identify three classes: (1) follow Democratic candidates only, (2) follow candidates from both parties, and (3) follow Republican candidates only. We use the class $c$ as the dependent variable and formulate the probability of each observation in a multinomial logistic setting [24]:

$$P1 = Pr(c=1) = \frac{e^{\mathbf{x}\beta_1}}{e^{\mathbf{x}\beta_1} + 1 + e^{\mathbf{x}\beta_3}}$$

$$P2 = Pr(c=2) = \frac{1}{e^{\mathbf{x}\beta_1} + 1 + e^{\mathbf{x}\beta_3}}$$

$$P3 = Pr(c=3) = \frac{e^{\mathbf{x}\beta_3}}{e^{\mathbf{x}\beta_1} + 1 + e^{\mathbf{x}\beta_3}}$$

where **x** is the vector of explanatory variables: number of posted tweets, number of followers, being a journalist (binary), gender, celebrity and year controls. Notice that the coefficients for the second class (following candidates from both parties) have been normalized to 0 to solve the identification problem.

The log-likelihood function then takes the form:

$$\ln L = \sum_{i=1}^{n}[\delta_{1i}\ln(P1) + \delta_{2i}\ln(P2) + \delta_{3i}\ln(P3)]$$

where $\delta_{ij}=1$ if i=j and 0 otherwise. Note that logistic regression, which we will use to differentiate the celebrity effects on Hillary Clinton and Bernie Sanders, is a special case of the multinomial logistic regression with only two choices instead of three.

## Results

In this section, we report on (1) election *follow* patterns observed on Twitter (2) negative binomial regression analysis of the number of candidates that one follows (3) multinomial logistic regression analysis of the effects of celebrity on the choice of candidates and (4) further logistic regression analysis between Hillary Clinton and Bernie Sanders.

*Election Follow Patterns on Twitter*

In Table V, we report on how committed each candidate's followers are. By commitment, we mean how many of the followers follow only that one specific candidate. It can be seen that Clinton, Trump and Sanders have highest percentages of 'committed' followers in the Twitter sphere, whereas only 9 percent of Bush's 529,820 followers follow him alone and 89 percent of Cruz's 1,012,955 followers follow other candidates besides Cruz. This suggests that while having a large number of followers is always beneficial, not all followers are equally committed.

To overcome this problem, we propose a simple and intuitive method to weight each follower by the reciprocal of the total number of candidates that he or she is following. For example, an individual who follows Bernie Sanders, Donald Trump and Ted Cruz will receive a weight of $\frac{1}{3}$, and an individual who follows Hillary Clinton only will receive a weight of 1. Mathematically, the Twitter share of candidate $j$ is then calculated as:

$$\text{share}_j = \frac{\sum_{i=1}^{n} \delta_{ij}\text{weight}_i}{\sum_{k=1}^{m}\sum_{i=1}^{n} \delta_{ik}\text{weight}_i}$$
$$\text{weight}_i = \frac{1}{\sum_{k=1}^{m} \delta_{ik}}$$

Table V
FOLLOWER ENGAGEMENT FOR EACH CANDIDATE (IN DECIMALS)

| Candidate | # 1 | # 2 | # 3 | # 4 | # 5+ |
|---|---|---|---|---|---|
| Chafee | 0.39 | 0.13 | 0.08 | 0.06 | 0.34 |
| Clinton | 0.75 | 0.16 | 0.04 | 0.02 | 0.04 |
| O'Malley | 0.29 | 0.23 | 0.17 | 0.07 | 0.24 |
| Sanders | 0.6 | 0.23 | 0.07 | 0.03 | 0.07 |
| Webb | 0.15 | 0.13 | 0.1 | 0.09 | 0.52 |
| Bush | 0.09 | 0.16 | 0.14 | 0.11 | 0.51 |
| Carson | 0.24 | 0.19 | 0.14 | 0.12 | 0.3 |
| Christie | 0.11 | 0.13 | 0.12 | 0.09 | 0.56 |
| Cruz | 0.11 | 0.17 | 0.18 | 0.15 | 0.39 |
| Fiorina | 0.41 | 0.15 | 0.08 | 0.07 | 0.29 |
| Kasich | 0.25 | 0.13 | 0.1 | 0.08 | 0.44 |
| Huckabee | 0.28 | 0.15 | 0.11 | 0.09 | 0.37 |
| Paul | 0.2 | 0.16 | 0.14 | 0.12 | 0.38 |
| Rubio | 0.25 | 0.16 | 0.15 | 0.13 | 0.32 |
| Trump | 0.72 | 0.14 | 0.05 | 0.03 | 0.06 |
| Walker | 0.14 | 0.08 | 0.08 | 0.08 | 0.62 |

Note: '#5+' stands for 'following five or more presidential candidates. For example, six percent of Trump followers follow five or more candidates.

where n is the total number of followers (15,455,122), m is the total number of candidates (16), $\delta_{ik}$ is 1 if individual i follows candidate k and 0 otherwise.

After applying this weighting mechanism, we find the Twitter share of the leading candidates, such as Donald Trump, Hillary Clinton and Bernie Sanders, further increases. Their aggregated share of Twitter followers rises from 68.7% to 80.1% (Figure 3).

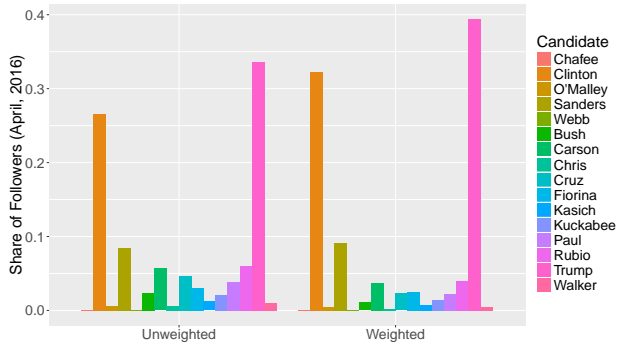

Figure 4. Share of the three leading candidates Trump, Clinton and Sanders further increases after weighting the followers.

We further analyze the top 15 most frequent patterns present in the Twitter sphere (Table VI). One immediate observation is that Trump, Clinton and Sanders are the three dominant forces in the Twitter sphere. 34.5% of the individuals recorded in our exhaustive dataset are following Donald Trump alone. 28.4% are following Hillary Clinton alone. 7.2% are following Sanders alone. These three groups account for 69.9% of the entire recorded population in our dataset. Individuals who follow only Marco Rubio or Carson or Fiorina make up no more than 2 percent of the population. Individuals who follow both Clinton and Trump constitute 3 percent of the entire recorded population. This number is surprisingly low and suggests that Twitter *follow* behavior is more of a signal of support (or interest) than mere communication as far as the presidential campaign is concerned. Other frequent 2-itemsets [13] include Carson and Trump (1%), Sanders and Trump 0.6/% and Rubio and Trump (1%). The only 3-itemset among the top 15 frequent patterns is Clinton, Sanders and Trump (0.5%).

Table VI
TOP 15 MOST FREQUENT ITEMS IN THE ELECTION'S TWITTER SPHERE

| 1 | 0.345 | Trump |
|---|---|---|
| 2 | 0.284 | Clinton |
| 3 | 0.072 | Sanders |
| 4 | 0.030 | Clinton Trump [2-itemset] |
| 5 | 0.021 | Rubio |
| 6 | 0.021 | Clinton Sanders [2-itemset] |
| 7 | 0.020 | Carson |
| 8 | 0.018 | Fiorina |
| 9 | 0.011 | Paul |
| 10 | 0.010 | Carson Trump [2-itemset] |
| 11 | 0.008 | Kuckabee |
| 12 | 0.007 | Cruz |
| 13 | 0.006 | Sanders Trump [2-itemset] |
| 14 | 0.006 | Rubio Trump [2-itemset] |
| 15 | 0.005 | Clinton Sanders Trump [3-itemset] |

We further examine how the decision of following one candidate correlates with that of following another candidate using the Pearson correlation coefficient. One immediate observation is that correlation between following candidates from the same party tends to be positive and correlation between following candidates from different parties tends to be negative (Figure 5). In particular, the correlation is -0.51 between Clinton and Trump and -0.22 between Sanders and Trump. By contrast, Marco Rubio and Ted Cruz have a strong and positive correlation coefficient of 0.43. This constitutes our first piece of evidence that individuals on Twitter are also polarized [6].

Motivated by the fact that Twitter *follow* behavior appears to cluster around the two parties, we refer to individuals who follow Democratic candidates exclusively as *Democrat followers* and refer to individuals who follow Republican candidates exclusively as *Republican followers* and lastly we refer to those who follow candidates from both parties as *Independent followers*. Note that this definition is based on Twitter *follow* behavior not on real party affiliation. It turns out that 92% of the 15.5 million followers are either Democrat followers or Republican followers, i.e., they are following candidates from only one party not both parties, which lends further support to the idea that the public are polarized on Twitter [6].

*Follow the Candidates*

Having summarized the election $follow$ patterns as a whole, we are now ready to analyze the factors behind an individual's decision to follow a certain number of candidates. In particular, while the marginal cost of following an extra candidate is close to zero, most individuals choose to follow only 1 or 2 candidates (Figure 5).

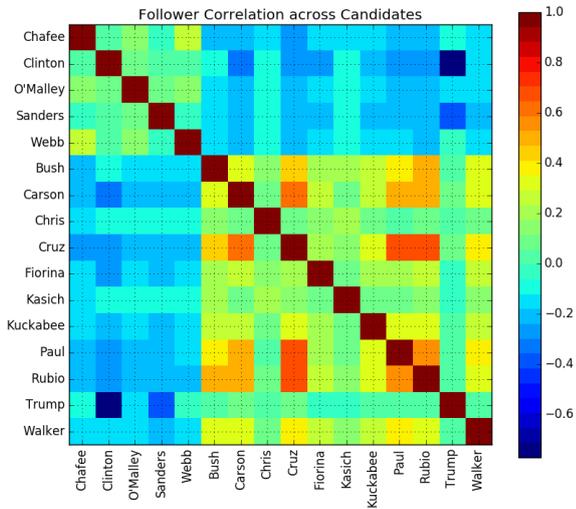

Figure 5. Party clustering observed in Twitter following behavior. Individuals who follow Trump are more likely to follow Ted Cruz and Marco Rubio and less likely to follow Hillary Clinton or Bernie Sanders.

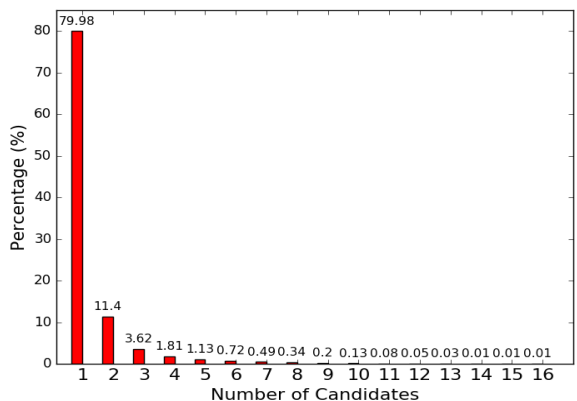

Figure 6. This figure is generated using all 15.5 million observations. In spite of the low marginal cost of following extra people on Twitter, most individuals chose to follow no more than 2 candidates during the 2016 U.S. presidential election.

In the regression, we use the number of candidates that one follows (# candidates) as the dependent variable. *social capital*, *journalist*, *gender* and *celebrity* are the four variables that we are particularly interested in. The coefficient on *social capital* would enable us to learn whether more prominent individuals tend to follow more candidates or not. The coefficient on *journalist* measures whether journalists tend to follow a larger number of the presidential candidates (and we expect the answer to be yes). *gender* measures the effects of being female. Following [5], [40], we expect the coefficient on *gender* to be negative, i.e., women tend to follow fewer candidates. The coefficient on celebrity tests whether celebrity followers behave differently from non-followers.

We report our regression results in Table VII. Across all the 7 specifications, we find that *tweets*, *social capital* and *journalist* are all positively correlated with the number of candidates that one chooses to follow and that *female* is negatively correlated with the number of candidates, which is consistent with [5], [40]. Most importantly, we observe that celebrity followers tend to follow a larger number of presidential candidates (Model 2 and Models 4-7). This does hold for Lady Gaga (Model 3), though, whose followers are more likely to follow a small number of candidates.

*Multinomial Logistic Regression: Celebrity Endorsements*

Having demonstrated that celebrity followers behave differently from non-followers in the number of candidates that they choose to follow, in this subsection we analyze the alignment and misalignment between celebrities and celebrity followers in political preferences. In particular, we use three-class multinomial logistic regression to examine whether followers of Lady Gaga and Lebron James, both of whom have explicitly endorsed Hillary Clinton, thus revealing support for Democratic causes, are more likely to follow Democratic candidates exclusively. Similarly, we study the preferences of the individuals who follow Willie Robertson and Toby Keith, and the preferences of those who follow Cornel West and Susan Sarandon.

We report our results in Table VIII. Using *Independent* as the baseline for comparison, we first examine the effects of posted tweets, social capital and occupation. Across all specifications, we find that people who have posted a large number of tweets, people with high social capital and people working as a journalist are more likely to be *Independent* followers, i.e., following candidates from both parties. We also observe that women are more likely to follow Democrats than men, as indicated by the positive coefficient on female for *Democrat* and negative coefficient for *Republican*.

From Columns 2 to 7, we examine the role of the six different celebrities in determining which candidates to follow. With the exception of Willie Robertson for *Republican*, the celebrity coefficients are all negative. To calculate the differences in predicted probability of political preference between celebrity followers and non-followers, we use the equations introduced in the section on Materials and Methods, taking all other variables at mean value while varying the binary variable celebrity from 0 to 1:

$$P1 = Pr(c=1) = \frac{e^{\mathbf{x}\beta_1}}{e^{\mathbf{x}\beta_1} + 1 + e^{\mathbf{x}\beta_3}}$$

TABLE VII
NEGATIVE BINOMIAL REGRESSION OF THE NUMBER OF CANDIDATES

|  | Model 1 Baseline | Model 2 James | Model 3 Gaga | Model 4 Willie | Model 5 Toby | Model 6 Cornel | Model 7 Susan |
|---|---|---|---|---|---|---|---|
| Tweets | 2.427*** (0.105) | 2.419*** (0.106) | 2.443*** (0.105) | 2.428*** (0.105) | 2.401*** (0.105) | 2.309*** (0.106) | 2.341*** (0.106) |
| Social Capital | 1.282*** (0.298) | 1.277*** (0.299) | 1.322*** (0.297) | 1.252*** (0.300) | 1.256*** (0.300) | 1.203*** (0.302) | 1.229*** (0.303) |
| Journalist | 0.201*** (0.0181) | 0.203*** (0.0181) | 0.200*** (0.0181) | 0.206*** (0.0181) | 0.207*** (0.0181) | 0.191*** (0.0181) | 0.199*** (0.0181) |
| Female | -0.0536*** (0.00230) | -0.0499*** (0.00231) | -0.0512*** (0.00230) | -0.0548*** (0.00230) | -0.0570*** (0.00230) | -0.0534*** (0.00230) | -0.0560*** (0.00230) |
| Lebron James |  | 0.0441*** (0.00310) |  |  |  |  |  |
| Lady Gaga |  |  | -0.0519*** (0.00296) |  |  |  |  |
| Willie |  |  |  | 0.480*** (0.00847) |  |  |  |
| Toby |  |  |  |  | 0.395*** (0.00796) |  |  |
| Cornel |  |  |  |  |  | 0.256*** (0.00746) |  |
| Susan |  |  |  |  |  |  | 0.202*** (0.00860) |
| Constant | 0.299** (0.108) | 0.298** (0.108) | 0.304** (0.108) | 0.301** (0.107) | 0.302** (0.107) | 0.291** (0.107) | 0.306** (0.107) |
| lnalpha |  |  |  |  |  |  |  |
| Constant | -4.320*** (0.0600) | -4.328*** (0.0604) | -4.330*** (0.0605) | -4.442*** (0.0673) | -4.414*** (0.0655) | -4.364*** (0.0625) | -4.341*** (0.0611) |
| Observations | 557777 | 557777 | 557777 | 557777 | 557777 | 557777 | 557777 |

Standard errors in parentheses
* $p < 0.05$, ** $p < 0.01$, *** $p < 0.001$

$$P2 = Pr(c=2) = \frac{1}{e^{\mathbf{x}\beta_1} + 1 + e^{\mathbf{x}\beta_3}}$$

$$P3 = Pr(c=3) = \frac{e^{\mathbf{x}\beta_3}}{e^{\mathbf{x}\beta_1} + 1 + e^{\mathbf{x}\beta_3}}$$

We report the calculated probabilities in Figure 7. It can be seen that even though Lebron James himself has endorsed Hillary Clinton, the probability of Lebron James's followers following only Democratic candidates is actually lower than that of non-followers. This also holds for Lady Gaga. What is special for Lady Gaga is that her followers are actually more likely to be following Republican candidates exclusively than non-followers. For both James and Lady Gaga, it is clear that there exists misalignment in political preferences with their followers.

The findings for Donald Trump's endorsers and for Bernie Sanders' endorses, by contrast, are consistent with what existing literature would predict. Individuals who follow Willie Robertson and Toby Keith (middle row) are more likely to be Republican followers and less likely to be Democrat followers. Compared with non-followers, those who follow Cornel West and Susan Sarandon (bottom row) are more likely to be Democrats and less likely to be Republican followers. For the latter four celebrities, we observe alignment of political preference between the celebrities and celebrity followers.

*Logistic Regression: Intra-party Celebrity Effects*

When comparing between *Democratic followers, Republican followers* and *Independent followers*, we observe strong alignment between Sanders' endorsers and their followers and misalignment between Clinton's endorsers and their followers. In this subsection, we restrict our comparison to Democratic followers who either follow Clinton or Sanders but not both, which constitute 36% of our observations. In particular, we will analyze (1) whether following celebrities who have explicitly endorsed Hillary Clinton can increase the probability of a Democrat follower following Hillary Clinton over Sanders and (2) whether following celebrities who have explicitly endorsed Bernie Sanders can decrease the probability of a Democratic follower following Hillary Clinton. The dependent variable, which is binary, is 1 if and only if the individual follows Hillary Clinton.

We report our results in Table IX. From Columns 1 and 2, we observe that individuals who follow Lebron James and Lady Gaga are more likely to be Clinton followers. By contrast, those who follow Cornel West and Susan Sarandon are less likely to follow Clinton, which suggests that Cornel's and Sarandon's followers are more likely to be following Sanders instead. One implication is that during primaries when candidates from the same party compete against each other, celebrities can already start to play an important role campaigning for the candidate they support.

DISCUSSION

Celebrities, if they so intend, could leverage their star power to affect the course of political and public life. In addition to the 2016 U.S. presidential election, Kevin Durant's

Table VIII
3-CLASS MULTINOMIAL ANALYSIS

|  | Model 1 Baseline | Model 2 James | Model 3 Gaga | Model 4 Willie | Model 5 Toby | Model 6 Cornel | Model 7 Susan |
|---|---|---|---|---|---|---|---|
| **Democrat** | | | | | | | |
| Tweets | -0.384 (0.446) | -0.296 (0.450) | -0.333 (0.447) | -0.336 (0.448) | -0.237 (0.450) | -0.159 (0.452) | -0.123 (0.452) |
| Social Capital | -7.334** (2.579) | -7.145** (2.593) | -6.831** (2.543) | -6.990** (2.530) | -6.937** (2.520) | -7.119** (2.591) | -6.540** (2.515) |
| Journalist | -0.640*** (0.0608) | -0.668*** (0.0610) | -0.651*** (0.0609) | -0.649*** (0.0608) | -0.655*** (0.0608) | -0.629*** (0.0609) | -0.640*** (0.0609) |
| Female | 0.490*** (0.0106) | 0.438*** (0.0107) | 0.506*** (0.0106) | 0.494*** (0.0106) | 0.501*** (0.0106) | 0.487*** (0.0106) | 0.500*** (0.0106) |
| Lebron James | | -0.594*** (0.0131) | | | | | |
| Lady Gaga | | | -0.302*** (0.0121) | | | | |
| Willie | | | | -1.484*** (0.0537) | | | |
| Toby | | | | | -1.412*** (0.0432) | | |
| Cornel | | | | | | -0.457*** (0.0261) | |
| Susan | | | | | | | -0.543*** (0.0288) |
| Constant | 1.588*** (0.413) | 1.619*** (0.413) | 1.627*** (0.413) | 1.586*** (0.413) | 1.583*** (0.413) | 1.627*** (0.413) | 1.581*** (0.413) |
| **Independent** | | | | | | | |
| **Republican** | | | | | | | |
| Tweets | -22.70*** (0.618) | -22.62*** (0.619) | -22.66*** (0.619) | -22.68*** (0.618) | -22.63*** (0.619) | -21.60*** (0.619) | -21.77*** (0.620) |
| Social Capital | -29.51*** (4.992) | -29.51*** (5.000) | -27.82*** (4.918) | -30.19*** (5.048) | -30.05*** (5.029) | -26.38*** (4.822) | -25.17*** (4.828) |
| Journalist | -1.838*** (0.0716) | -1.859*** (0.0717) | -1.858*** (0.0717) | -1.836*** (0.0716) | -1.842*** (0.0716) | -1.812*** (0.0721) | -1.840*** (0.0718) |
| Female | -0.0518*** (0.0104) | -0.0895*** (0.0104) | -0.0237* (0.0104) | -0.0528*** (0.0104) | -0.0493*** (0.0104) | -0.0582*** (0.0104) | -0.0304** (0.0104) |
| Lebron James | | -0.406*** (0.0123) | | | | | |
| Lady Gaga | | | -0.577*** (0.0118) | | | | |
| Willie | | | | 0.179*** (0.0420) | | | |
| Toby | | | | | -0.206*** (0.0348) | | |
| Cornel | | | | | | -2.028*** (0.0323) | |
| Susan | | | | | | | -2.178*** (0.0374) |
| Constant | 1.203** (0.449) | 1.226** (0.449) | 1.265** (0.450) | 1.203** (0.449) | 1.201** (0.449) | 1.300** (0.452) | 1.189** (0.448) |
| Observations | 557777 | 557777 | 557777 | 557777 | 557777 | 557777 | 557777 |

Standard errors in parentheses
* $p < 0.05$, ** $p < 0.01$, *** $p < 0.001$

refusal to visit the White House [2] and Lady Gaga's meeting with Dalai Lama [3] both sent strong political messages too. Our paper demonstrates the political preference of a celebrity could acutally differ systematically from that of the fan base. In future research it would be beneficial to evaluate how celebrity endorsements affect the political behavior of individuals.

[2] http://ftw.usatoday.com/2017/08/kevin-durant-white-house-warriors-trump.
[3] https://www.theguardian.com/music/2016/jun/28/china-lady-gaga-ban-list-hostile-foreign-forces-meeting-dalai-lama.

CONCLUSION

This paper studies the online behavior of celebrity followers during the 2016 U.S. presidential campaign and analyzes the alignment and misalignment of political preferences between celebrities and celebrity followers. Building from an exhaustive dataset that includes 15.5 million records, taking advantage of three information channels (name, image, description), and applying various regression models, we (1) explored the frequent patterns of the 2016 U.S. presidential campaign on Twitter, calculated the weighted presence for each candidate, and measured the extent to which individu-

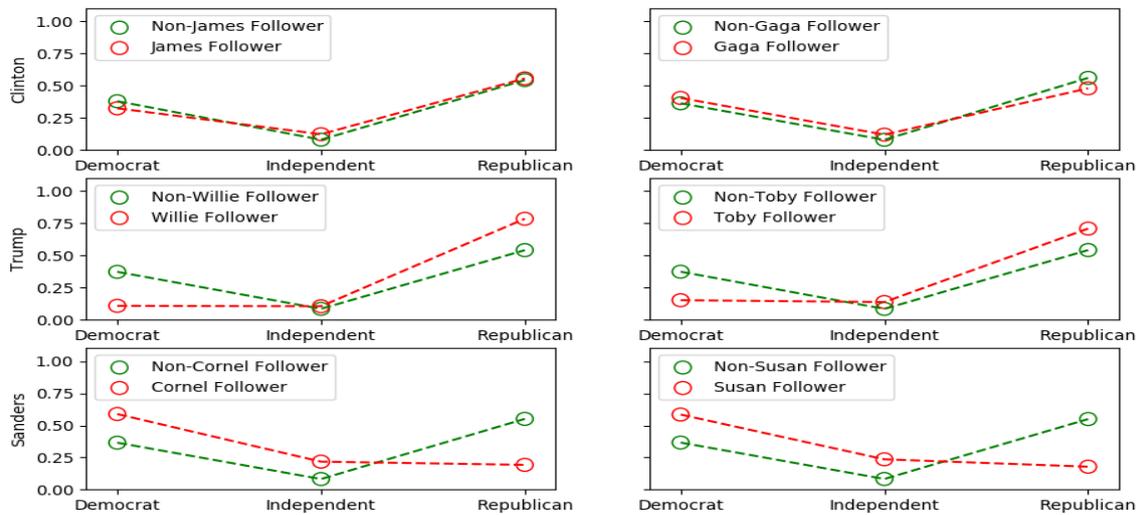

Figure 7. Misalignment is observed between Hillary Clinton's endorsers and their followers (top row). Alignment is observed between Donald Trump's endorsers (middle row), Bernie Sanders endorsers (bottom row) and their followers.

Table IX
INTRA-PARTY CELEBRITY EFFECTS: A LOGISTIC REGRESSSION

| Explanatory Variable | Model 1 Lebron James | Model 1 Lady Gaga | Model 3 Cornel | Model 4 Susan |
|---|---|---|---|---|
| Social Capital | 5.358 | 3.963 | 5.491 | 5.223 |
|  | (5.102) | (4.793) | (5.119) | (5.120) |
| Journalist | 1.089*** | 1.088*** | 1.064*** | 1.058*** |
|  | (0.118) | (0.118) | (0.118) | (0.118) |
| Female | 0.124*** | 0.0641*** | 0.0748*** | 0.0791*** |
|  | (0.0113) | (0.0113) | (0.0112) | (0.0112) |
| Lebron James | 0.909*** |  |  |  |
|  | (0.0214) |  |  |  |
| Lady Gaga |  | 0.977*** |  |  |
|  |  | (0.0173) |  |  |
| Cornel |  |  | -0.487*** |  |
|  |  |  | (0.0325) |  |
| Susan |  |  |  | -0.185*** |
|  |  |  |  | (0.0367) |
| Constant | 0.702 | 0.616 | 0.767* | 0.716 |
|  | (0.384) | (0.387) | (0.385) | (0.384) |
| Observations | 190187 | 190187 | 190187 | 190187 |

Standard errors in parentheses
* $p < 0.05$, ** $p < 0.01$, *** $p < 0.001$

als on Twitter are polarized, (2) studied how following a celebrity has had an effect on the number of candidates that one chooses to follow, (3) found alignment between Trump's endorsers and their followers, alignment between Sanders' endorsers and their followers, and misalignment between Clinton's endorsers and their followers, (4) found that when considering Democrat followers only, followers of Clinton's endorsers tend to favor Hillary Clinton over Bernie Sanders, and followers of Sanders' endorsers tend to favor Sanders over Clinton.